\documentclass[twocolumn,aps,pra,showpacs,nofootinbib,groupedaddress]{revtex4-1}
\usepackage{epsfig,amssymb,amsmath,mathrsfs,color,comment} 
\begin{document}

\title{Introduction to the book ``Quantum Theory: Informational Foundations and Foils"}
\author{Giulio Chiribella} 
\affiliation{Department of Computer Science, University of Oxford, Wolfson Building, Parks Road, Oxford, UK}
\affiliation{Canadian Institute for Advanced Research,
CIFAR Program in Quantum Information Science, Toronto, ON M5G 1Z8}
\affiliation{Department of Computer Science, The University of Hong Kong, Pokfulam Road, Hong Kong}
\author{Robert W. Spekkens}
\affiliation{Perimeter Institute for Theoretical Physics, 31 Caroline Street North, Ontario,  Ontario N2L 2Y5, Canada.}

\begin{abstract}
We present here our introduction to the contributed volume  ``Quantum Theory: Informational Foundations and Foils", Springer Netherlands (2016).   
It highlights recent trends in quantum foundations and offers  an overview of the contributions appearing in the book.  

\end{abstract}
\maketitle  

This paper contains the introduction to the contributed volume  ``Quantum Theory: Informational Foundations and Foils", Springer Netherlands (2016).   
The table of contents of the book, with links to  the arXiv versions of the other chapters, is as follows. 

\bigskip 

\noindent {\bf Part I:  \\ Foil Theories}
\begin{enumerate}
\item[] {\bf Chapter 1\\ Optimal Information Transfer and\\ Real-Vector-Space Quantum Theory}  \\  
William K. Wootters, \\
https://arxiv.org/abs/1301.2018   
\item[] {\bf Chapter 2\\ Almost Quantum Theory}  \\  
Benjamin Schumacher and Michael D. Westmoreland \\
https://arxiv.org/abs/1204.0701
\item[] {\bf Chapter 3\\ Quasi-Quantization:~Classical~Statistical Theories with an Epistemic Restriction}  \\ 
Robert W. Spekkens\\
https://arxiv.org/abs/1409.5041
\end{enumerate}

\bigskip 

\noindent {\bf Part II:  \\ Axiomatizations}
\begin{enumerate}
\item[] {\bf Chapter 4\\ 
Information-Theoretic Postulates for Quantum Theory}  \\ 
Markus P. M\"uller and Lluis Masanes\\
https://arxiv.org/abs/1203.4516
\item[] {\bf Chapter 5\\ 
Quantum From Principles}  \\ 
Giulio Chiribella, Giacomo M. D'Ariano, and Paolo Perinotti\\
https://arxiv.org/abs/1506.00398
\item[] {\bf Chapter 6\\
Reconstructing Quantum Theory}\\
Lucien Hardy\\
https://arxiv.org/abs/1303.1538
\item[] {\bf Chapter 7\\
The Classical Limit of a Physical Theory and the Dimensionality of Space}\\
Borivoje Dakic and  \v Caslav Brukner\\
https://arxiv.org/abs/1307.3984
\item[] {\bf Chapter 8\\
Some Negative Remarks on Operational Approaches to Quantum Theory
}\\
Christopher A. Fuchs and Blake C. Stacey\\
https://arxiv.org/abs/1401.7254
\end{enumerate}

\bigskip 

\noindent {\bf Part III: \\ Categories and Convex Sets}
\begin{enumerate}
\item[] {\bf Chapter 9\\ 
Generalised Compositional Theories \\ and Diagrammatic Reasoning }  \\ 
Bob Coecke, Ross Duncan, Aleks Kissinger, and Quanlong Wang\\
https://arxiv.org/abs/1506.03632
\item[] {\bf Chapter 10\\ 
Post-Classical Probability Theory}  \\ 
Howard Barnum and Alexander Wilce\\
https://arxiv.org/abs/1205.3833
\end{enumerate}

\bigskip 

\noindent {\bf Part IV: \\ Quantum Versus Super-Quantum Correlations}
\begin{enumerate}
\item[] {\bf Chapter 11\\ 
Information Causality}  \\ 
Marcin Paw\l owski and Valerio Scarani\\
https://arxiv.org/abs/1112.1142
\item[] {\bf Chapter 12\\ 
Macroscopic Locality}  \\ 
Miguel Navascu\'es  \\
(arXiv version  not available at the moment)
\item[] {\bf Chapter 13\\ 
Guess Your Neighbor's Input: No Quantum Advantage but an Advantage for
                               Quantum Theory}  \\ 
Antonio Ac\'in, Mafalda L. Almeida, Remigiusz Augusiak, Nicolas Brunner\\
https://arxiv.org/abs/1205.3076
\item[] {\bf Chapter 14\\ 
 The Completeness of Quantum Theory for Predicting Measurement Outcomes}\\
Roger Colbeck and Renato Renner  \\ 
https://arxiv.org/abs/1208.4123
\end{enumerate}

\noindent

\vspace{50pt}

\noindent {\large \bf Introduction}

\medskip 

The foundations of Quantum Mechanics are 
experiencing a golden age. In a timespan of less than two decades, 
an astonishing number of new results, ideas, and frameworks have 
revolutionized the way we think about the subject.   A new research community is emerging worldwide, attracting scientists from a diverse spectrum of disciplines including  physics, computer science, and mathematics. 
The keyword ``foundations"  
is now included in the strategic priorities of many research institutions and funding agencies, and it regularly features as  one of the hot topics highlighted in international conferences.   

The abundance of ideas, approaches, and resources that have emerged poses some challenges however.  For one, having a global vision of the field and reflecting on its high level goals is becoming increasingly difficult.  For another, the sheer number of different frameworks that have been put forward 
risks creating a tower of Babel effect, fragmenting the community into smaller cliques
that are unable to talk  to one another. In addition, researchers who are joining the field   have to cope with a fast-moving landscape where it can be hard to identify stable reference points.   

These considerations led us to the project of this book, which aims to showcase
 the state of the art  in quantum foundations.    
The book provides a collection of articles that deal with influential ideas in the field today, 
revealing the diversity of approaches on the one hand, and highlighting 
the common threads among them on the other.

\section{Characteristics of the new wave of quantum foundations}

We start by   outlining what is distinctive about the foundational research   that this book aims to portray.   
\subsection{A pragmatic perspective}

It is useful to distinguish between what one might call \emph{dynamicist} and \emph{pragmatist}  traditions in physics.  Within the dynamicist tradition, the physicist's job is to describe the natural dynamical behaviour of a system, without reference to human agents or their purposes.   In the pragmatist's approach, on the other hand, the laws of physics are characterized  in terms of the extent to which we can learn and control the behaviour of physical systems.
The distinction between the dynamicist and pragmatist points of view is nicely represented in competing formulations of the second law of thermodynamics.
One that is clearly in the dynamicist tradition is Clausius's original statement: 
\begin{quote}
{\em Heat can never pass from a colder to a warmer body without some other change, connected therewith, occurring at the same time \cite{clausius}.}
\end{quote}
On the other hand, the version of the Kelvin-Planck statement that is found in most textbooks is clearly pragmatic: 
\begin{quote}
{\em It is impossible to devise a cyclically operating device, the sole effect of which is to absorb energy in the form of heat from a single thermal reservoir and to deliver an equivalent amount of work \cite{rao}. }
\end{quote}

Quantum theory has always partaken in both traditions.    Indeed, Schr\"{o}dinger's wave mechanics and Heisenberg's matrix mechanics were distinguished in part by the fact that Schr\"{o}dinger, following de Broglie's lead, sought to provide a description of the motion of particles, while Heisenberg, following Bohr's lead, espoused an operational philosophy and took his formalism to merely describe what would be observed in certain experimental circumstances.    The new foundational work represents a renewed interest in exploring quantum theory within the pragmatist tradition.     

\subsection{Quantum foundations in the light of quantum information}

 The newfound popularity of the pragmatist tradition is 
tightly  connected with 
the rise of quantum information theory.  
The real innovation of the recent foundational work  is in the way researchers conceive the difference between quantum and classical theories \cite{hardyspek}.  Historically, quantum theory was taken to consist entirely of \emph{restrictions} on our information-gathering ability; 
 think, for instance, of the restriction imposed by the uncertainty principle.  The quantum information revolution overturned this notion: a quantum world in fact holds new possibilities for information-processing tasks---in particular, communication tasks, cryptographic tasks, and computational tasks---that could not be accomplished  in classical physics. 

 
 
Milestone applications of quantum information, such as secure quantum key distribution~\cite{BB84,ekert}, ultrafast quantum algorithms~\cite{shor,grover}, teleportation \cite{teleportation}, and dense coding~\cite{densecoding}, stimulated the imagination of quantum theorists, and led them to ask questions that moved beyond the usual topics of foundational discussions:  
\emph{Which principles of quantum theory can account for its information-processing advantages?
Does the possibility of achieving one kind of information-processing advantage imply the possibility of achieving others?  Is quantum theory the \emph{only} theory where these advantages arise?}  These questions were at the center of 
an influential research programme, launched by  Fuchs \cite{fuchs,fuchs1} and Brassard \cite{brassard}, that aimed to understand quantum theory in the light of quantum information.  
More specifically, the idea was to take certain facts about the information-processing features of a quantum world, for instance, the possibility of secure key distribution and the impossibility of secure bit commitment, and derive the quantum formalism from these.   This line of inquiry gave birth to a new breed of foundational research with more pragmatic ambitions, with practitioners that split their time between developing novel practical applications of quantum information and achieving a deeper foundational understanding of quantum theory, with each activity informing the other.

\subsection{The shift from interpretation  to reconstruction}


Traditionally, the focus of many quantum foundations researchers was the \emph{interpretation} of quantum theory.  In most such works,  the formalism of quantum theory was taken as given, and the goal was to infer from this formalism the correct story to tell about the nature of reality---typically, a story of dynamicist flavour.    
The Everett interpretation~\cite{everett} and the deBroglie-Bohm interpretation~\cite{bohm} are examples.  
Models incorporating physical collapses~\cite{grw,gpr}
are also proposed 
in an effort to secure a dynamicist story about quantum theory.     

By contrast, the focus of the new wave is  the \emph{reconstruction} of quantum theory from physical principles.   Contemporary researchers are looking for an answer to Wheeler's famous question \emph{``Why the quantum?"} \cite{wheeler} and  are driven to 
understand the origin of the  formalism itself.   
Textbook postulates such as  ``a physical system is described by a complex Hilbert space", ``pure states are described by  unit vectors", ``outcome probabilities are given by the Born rule", and ``systems combine by the tensor product rule'' are now regarded as abstract mathematical statements in need of a more fundamental explanation.  Such an explanation would be akin in spirit to Einstein's  derivation of the Lorentz transformations from the light postulate and the principle of relativity.






The goal is to find 
a compelling set of axioms that singles out quantum theory from among all possible theories.  
Finding an appealing axiomatization is a problem that has a long tradition, starting with the work of Birkhoff and von Neumann~\cite{birkvon} and continuing through the works of Mackey~\cite{mackey}, Ludwig~\cite{ludwig}, Piron \cite{piron}, and the field of quantum logic~\cite{qlogic1,qlogic2}.  What distinguishes the axiomatic work being pursued today 
is the use of notions inspired by quantum information theory, the emphasis on composite systems, the focus on finite-dimensional Hilbert spaces, and an insistence on axioms that are operationally meaningful.  


\subsection{The operational framework}
Any question of the form ``why \emph{this}?'' is implicitly asking ``why not \emph{that}?''.  Therefore, to tackle Wheeler's question,
one first of all needs to be able to conceive of alternatives to quantum theory, ways the world \emph{might have been}.  
In short, one requires a framework for describing a broad range of physical theories, including quantum and classical theories, but allowing 
more exotic alternatives as well.  

 One way to achieve such a framework  is to focus on a strictly operational formulation of physical theories.
An operational formulation is one wherein the primitive concepts are preparation procedures, transformation procedures, and measurement procedures, each understood as a specification of a list of instructions for an experimentalist, spelled out in sufficient detail that they could be implemented by any technician,  
 as with a good recipe.  The theory specifies a mathematical algorithm that fixes the probability distribution over outcomes for every possible measurement given every possible preparation and intervening transformation.  When phyical theories are operationally formulated, therefore, the only relevant differences between them are differences in the sorts of experimental statistics that they allow.

The operational approach encourages one to focus on a characterization of quantum theory in terms of experimental facts, and to consequently avoid, as much as is possible, making claims that go beyond what is strictly required to describe these facts. This sort of exercise can be very useful for freeing the mind from all the baggage of classical preconceptions and previous attempts to interpret the quantum formalism.
For many researchers, adopting this approach is not a rejection of the need for providing a dynamicist account of quantum theory, nor is it necessarily an endorsement of the notion that a physical theory is \emph{nothing more} than an algorithm for predicting experimental statistics.  Rather, it is considered an effective methodological tool for making progress on questions about the origin of the quantum formalism.










\subsection{Foil theories}

A distinctive characteristic of contemporary foundations is the exploration of alternatives to quantum theory, that is, \emph{foil theories}. A foil to $X$ is something that helps to highlight the distinctive characteristics of $X$ by contrasting with it~\footnote{
``Whenever I marry,'' she continued after a pause which none interrupted, ``I am resolved my husband shall not be a rival, but a foil to me.''  ---from Jane Eyre, by Charlotte Bront\"e.
}.  
Given a framework of possible theories that includes quantum theory, every nonquantum point in the landscape is a foil theory.  
Each such theory specifies a way the world might have been had it not been quantum.  

We use the term `foil' to highlight the attitude that is taken towards these theories: they are \emph{not} being proposed as empirical competitors to quantum theory, with grand ambitions of usurping its throne.
Rather, they serve to clarify what is distinctive about quantum theory.  For instance, if one can identify a foil theory that shares some set of features with quantum theory, then that set of features cannot possibly be 
a complete set of axioms for quantum theory.    Likewise,   constructing foil theories is an essential  step for proving the independence of a set of axioms: if one axiom is independent from another, then one should be able to devise a foil theory that satisfies the former but violates the latter. 

\subsection{Goals}
  
One of the ambitions of researchers in quantum foundations  is that
 the insights coming from their work will help  
 with some of the big challenges of contemporary physics,  
 such as the formulation of a quantum theory of gravity.  
Another ambition is to find alternatives to quantum theory that \emph{could}  eventually become empirical competitors.  Given an axiomatic derivation of quantum theory, one can consider modifying a single axiom in order to build a consistent alternative.  
Furthermore, this approach can be used to avoid an important pitfall of more ad hoc approaches to developing alternatives to quantum theory, namely, that the latter may inadvertently violate fundamental principles that one would prefer not to abandon.  A good example is the nonlinear modification of quantum theory proposed by Weinberg~\cite{weinberg} which was subsequently shown to allow for superluminal signalling~\cite{gisin} and also to violate the second law of thermodynamics for the normal definition of entropy~\cite{peres}.  In the axiomatic approach, the  fundamental principles that one wants to uphold can be built in from the outset. 


A more practical application of this foundational work is to advance quantum technologies.  Indeed, such work is beginning to clarify how information-processing capabilities can arise from foundational principles.  For instance, cryptography based on Bell-inequality violations~\cite{ekert,mayers} can be shown to be secure 
even if the devices used in the protocol are supplied by the adversary, as long as it is presumed that the adversary cannot signal superluminally~\cite{BHK,AGM}.    This idea, which originated from foundational works, led to an entire field of \emph{device-independent cryptography}  \cite{BHK,AGM,acin,maspir,vazivid}.

\section{Frameworks for operational theories}

It is worth  spending a few words on the specific 
frameworks
that have been developed in an attempt to achieve the aims described above. 
 Because existing frameworks were found insufficient, many researchers opted to construct a new canvas for their portrait of quantum theory, with quantum information processing serving as their muse. 
 The emphasis that is placed on the development of such frameworks is itself a distinctive trait of the new   wave of foundational research.



To the outsider, it is hard to appreciate the importance of constructing the framework.
But it is in fact a highly non-trivial task, where one is forced to make  fundamental choices as to what is considered ``general" (i.e. part of the notion of a physical theory)  and what is considered ``specific" (and hence a possible candidate for an axiom that identifies quantum theory).  What is at stake in the choice of a framework is the very definition of a physical theory.       
   
Note that having a framework for operational theories is not only useful as an instrument for axiomatizations,  but also as a playground for experimenting with alternative models of information processing.  Such frameworks are increasingly being used to attempt to describe nonclassical phenomena in a language that does not presume the correctness of quantum theory.  Not only is this pursued for the question of Bell inequality violations \cite{bellrmp,nonlocalityresource,prunit,distillation}, but also for a number of applications to computer science and physics, including the study of communication complexity \cite{vanDam,BrassardCommunicationComplexity}, non-local computation \cite{NonLocalComp},  measurement-based computation \cite{anders,raussendorf,duncanperdrix,horseman,rossmbqc},  games and interactive proof systems   \cite{raz,holenstein,ito1,ito2,kalai,rotem}, randomness amplification   \cite{rogerthesis,pironio,colbeck,free},  causal networks    \cite{fritz,henson,chaves},   computability \cite{islam}, complexity \cite{lee},  key distribution \cite{full},  bit commitment \cite{bitcommitmentharry,bitcommitment,CDP2010},   complementarity \cite{ross,harny}, no cloning \cite{nobroad,CDP2010,abramskyclon},  teleportation \cite{abracoecke,telep,CDP2010}, state discrimination \cite{statedist1,statedist2,statedist3},  entropy \cite{entropy1,entropy2,entropy3}, thermodynamics \cite{thermowehner,thermobrunner,thermoscandolo},  general resource theories \cite{resource}, and spacetime physics \cite{mueller3d,muellerblackhole}. 
This long list provides a good illustration of how fertile the development of new frameworks has been.  In the following, we identify the main directions along which the framework-building activity has developed so far.  
  

\subsection{The framework of convex operational theories}
 
 A particularly popular framework is that of {\em convex operational theories},
where  preparations, transformations, and measurements are represented by elements of suitable convex sets, the dimension of which is fixed by the nature of the physical systems involved in the experiment.    

The framework of convex operational theories is the contemporary descendant of the frameworks used in the tradition of operational quantum logic, in particular those introduced by Mackey \cite{mackey}, Ludwig \cite{ludwig}, and Davis and Lewis \cite{davieslewis}.    
In the new wave of quantum foundations, the first elaboration of this framework appeared
 in Hardy's 2001 axiomatization of quantum theory~\cite{hardy2001}.    With respect to earlier  works in quantum logic,  Hardy's framework distinguishes itself by being more manageable and intuitive, partly because of its focus on finite-dimensional systems.    
This approach was brought to  completion through a series of works by a number of other authors \cite{barrett,nobroad,telep,barnumwilce}.

\subsection{The category-theoretic framework}      

Due to the long tradition of using  convex sets to represent the state spaces of physical systems,    there is a strong temptation  to identify the operational approach with the 
framework of convex operational theories. 
 However, a substantial part of what defines a physical theory has nothing to do with convex sets, or even with probabilities. For example, operational notions such as composing two systems in parallel (this {\em and} that)  and composing two physical processes in a sequence (do this {\em and then} do that)
are more primitive than the notion of probability.   
Such notions of composition are the focus of the \emph{category-theoretic framework} initiated by Abramsky and Coecke \cite{abracoecke,cqm,guises,contemporarybob}. 
In this framework, the mathematical structure describing a general physical theory, in particular the two notions of composition and how they interact, is that of a strict symmetric monoidal category.
One of the characteristic features of the category-theoretic framework is that all the relations of interest can be encoded in diagrams, similar to those used in the representation of quantum circuits.    


 \subsection{The framework of operational-probabilistic theories}  

The lesson of the category-theoretic framework is that  the composition of systems and processes is fundamental to the operational structure of a theory and that one can talk about information processing without even having to mention probabilities.  On the other hand, the precise probabilistic predictions of an operational theory are sometimes a feature of interest.  
If one is interested in {\em both} the compositional and the probabilistic features of a theory, then the framework 
 of  \emph{operational-probabilistic theories}, recently developed by Chiribella, D'Ariano and Perinotti \cite{CDP2010,CDP2011,chiribella14} and Hardy \cite{hardy2010,hardy2011}, provides a  supplementation of the category-theoretic framework with probabilistic structure.

In this framework, the category-theoretic notions are used to define circuits of physical processes.  An experiment is represented by a closed circuit, starting from the preparation of a system and ending with a measurement having a particular outcome.  The probabilistic structure is added on top of the circuit framework by introducing a rule that assigns probabilities to these closed circuits. 
The result of this construction is that states, transformations, and measurements are represented by elements of suitable vector spaces, as they are in the framework of convex operational theories.  However, the framework of operational-probabilistic theories allows one to describe also theories where the state space is not convex, such as Spekkens' toy theory \cite{spekkens}. In addition, it allows one to treat causality as an emergent feature in a broader class of physical theories where causality is not assumed as part of the framework \cite{CDP2010}.

When we wish to refer to a framework that can describe features of experimental probabilities, while remaining noncommital about whether it is the framework of convex operational theories or the more general framework of operational-probabilistic theories, we shall speak simply of the framework of \emph{generalized probabilistic theories} (GPTs).  


 

\subsection{The device-independent framework}

Another popular framework is the \emph{device-independent framework}  \cite{poprohr,BHK,AGM,scaranireview}.   Here, an experiment is not parsed into preparations, transformations and measurements, with a physical system of a particular dimension acting as a causal mediary between these. Rather, the experiment is treated as a black box, characterized completely by how it maps classical inputs to classical outputs.  The roots of this approach can also be traced back to the quantum information revolution:   considering input-output black boxes is a natural   
approach to the design of  cryptographic protocols that are secure even if the functioning of the devices is not trusted.  
 In this context, proving the security of a  protocol independently of the inner workings of its   black box components is desirable because the components may have been designed by one's adversary. 

The device-independent framework is apt to capture the \emph{device-independent features} of quantum theory.  
 The paradigmatic example of a device-independent quantum feature is the Tsirelson bound \cite{tsirelsonbound}, which can be viewed as an upper bound on the probability that two cooperating players win a game, known as the \emph{CHSH} game after the seminal work of Clauser, Horne, Shimony and Holt \cite{chsh}). In the CHSH game, the inputs are the questions asked by a referee to the two players, and the outputs are their answers.  While playing the game, the players are allowed to share arbitrary  entangled states and are allowed to perform arbitrary local measurements on their systems. Still, their winning probability is upper bounded, independently of the states they prepare and of  the measurements they perform.  The bound is device-independent, in that it depends only on the validity of quantum theory.

The CHSH game is the problem that got the device-independent approach started, when Popescu and Rohrlich \cite{poprohr} and   Rastall \cite{rastall}  came up with a foil theory that is \emph{more nonlocal than quantum theory},~i.e.,~it guarantees to the players a higher winning probability in the CHSH game.   Nevertheless, any other game would define a device-independent feature of quantum theory.   
The ultimate device-independent feature   is the specification of the full set of correlations (i.e., the conditional probability of  the outputs given the inputs) that are achievable by local quantum measurements on a bipartite quantum state.  This is known as {\em the quantum set}.

A particularly active line of research in recent years has been the problem of \emph{deriving} device-independent features of quantum theory from information-theoretic principles. The ultimate dream of researchers working in this area is to derive the specific shape of the quantum set by using only device-independent axioms, that is, axioms that refer only to the conditional input-output probabilities.

Although  the study of information processing in generalized probabilistic theories and the study of device-independent features  have developed on separate tracks until now, the time is ripe for  uncovering connections between them. On the one hand, the tools developed in the study of axioms for generalized probabilistic theories may help to achieve a characterization of the quantum set, a project that is notoriously difficult.   On the other hand,  device-independent features may provide candidates for new axioms. A detailed discussion of the connections between the two frameworks
can be found in Ref. \cite{yuan2015}.

\section{Book synopsis} 
 
The information-theoretic characterization of quantum theory is a general direction that unites the efforts of the  new quantum foundationalists,  although below this umbrella there is an exceptional variety of different approaches and  goals.  The book aims to provide a panoramic view of the field, including some of the most promising directions that have emerged in the past decade.    It  is divided into four sections, corresponding to the following themes:  
\begin{enumerate}
\item Foil theories  (Chapters 1-3)
\item Axiomatizations (Chapters 4-8)
\item 
Categories and convex sets
(Chapters 9-10)
\item Quantum versus super-quantum correlations  (Chapters 11-14) 
\end{enumerate} 
This subdivision is meant as an aid for readers who are approaching the field for the first time and want to have an idea of the big picture.  
Many other organizational schemes would have worked just as well, and we therefore encourage readers to explore other paths through the various contributions.
In the following, we provide a  synopsis of the book through its four sections.  


\medskip 

 \subsection{Foil theories}    
   We open the book with three examples of foil theories.
   
   Wootters (chapter 1) considers {\em real quantum theory} \cite{stueck,araki,wootters}, which is the foil theory that results from replacing the complex field with the real field in the standard formalism of quantum theory.   He considers the information transfer from a preparation to a measurement and shows that for certain natural ways of quantifying this transfer---for instance, the mutual information between the angle of a polarizer that prepares a photon's polarization and the relative frequency of outcomes in a measurement of polarization---the information transfer is optimized for real quantum theory and not for complex quantum theory.  He further considers the question of whether some {\em other} notion of information transfer might pick out complex quantum theory rather than its real counterpart.

    Schumacher and Westmoreland (chapter 2) present {\em modal quantum theory} \cite{modal}, which replaces the complex field with a finite field.  This necessitates a more dramatic modification of the quantum formalism than is required to replace the complex field with the real field. The foil theory that they construct is {\em possibilistic} rather than {\em probabilistic}: it does not specify the probabilities of different measurement outcomes, but only which outcomes are possible and which are impossible.  Despite the fact that modal quantum theory is rather minimalist in the scope of states and measurements that it permits, it nonetheless reproduces a surprising number of qualitative features of quantum theory.

     Spekkens (chapter 3) considers a family of foil theories that arise from taking a classical statistical theory and imposing an epistemic restriction, that is, a restriction on the amount of knowledge any observer can have about the physical state of a classical system  \cite{spekkens}. Depending on the type of degree of freedom being considered, the resulting foil theory either describes a subset of the preparations, transformations and measurements allowed in the full quantum theory for that type of degree of freedom, or it describes a distortion of such a subset that is inequivalent in its predictions to quantum theory.  Both types are shown to reproduce a large number of phenomena that are usually taken to be distinctively quantum, but to lack others, thereby suggesting a distinction between weak and strong notions of nonclassicality.

\medskip 
   
\subsection{Axiomatizations}
       This part of the book   presents  three different axiomatizations of quantum theory (Chapters 4-6) along with two contributions on themes that are closely related to the axiomatic endeavour (Chapters 7-8).  
For reasons of space, all of the axiomatization chapters confine themselves to presenting an outline of the main ideas behind the derivation of the Hilbert space formalism, while omitting the technicalities that go into the mathematical derivations (these can be found, of course, by referring to the original research articles). 

Masanes and M\"uller (chapter 4) 
present their 2011 axiomatization of quantum theory~\cite{MM2011}.  We start our lineup of axiomatization here because this work is a  direct descendant of Hardy's seminal 2001 axiomatization, from which it inherits some of its axioms.  With respect to Hardy 2001, the main progress here is in the elimination of one axiom, called the ``Simplicity Axiom", which, compared to the others, seemed to be less motivated.   
Within both the Hardy 2001 and the Masanes-M\"uller 2011 axiomatizations, the feature that distinguishes quantum from classical theory is the fact that every two pure states are connected by a \emph{continuous} path of pure states.  

Chiribella, D'Ariano and Perinotti
 (chapter 5) present their  axiomatization \cite{CDP2011}.
The central axiom here is the Purification Postulate, stating that every mixed state of a given physical system can be  modelled as the marginal of a pure state of a larger composite system.    This requirement directly implies  many quantum features, such as no-cloning, teleportation, and the fact  that every irreversible process can be modelled as the result of a reversible interaction between the system and an environment that is subsequently discarded \cite{CDP2010}.  
A slogan for this axiomatization is that quantum theory is the only pure and reversible theory of information.

 
 We conclude our lineup of axiomatizations with Hardy (chapter 6) who presents his 2011 axiomatization~\cite{hardy2011}.  In this axiomatic scheme, the emphasis is on the perfect distinguishability of states and on the  possibility of performing computations reversibly.  Hardy proves that there  are the only two theories compatible with his new set of axioms: classical and quantum. Once this result is established, therefore, one can identify quantum theory by choosing any feature that distinguishes it from classical theory. Insofar as this work constitutes a significant development of Hardy's influential 2001 axiomatization and incorporates tools and ideas introduced by other authors working on axiomatization, it is a good illustration of the progress of the field in the last decade.
 
 Chapters 7 and 8 do not present new axiomatizations, but  nonetheless concern themselves with the axiomatization project.


Daki\'{c} and Brukner (chapter 7) note that within generalized probabilistic theories, experimental operations are described abstractly and do not make direct contact with more traditional concepts of physics, such as position in space, direction, and energy.  Their work aims to bridge this gap to some extent. They show that, within a suitable class of theories, quantum theory embedded in a three-dimensional space is the only theory satisfying the consistency requirement that every possible transformation of a single elementary system can be generated by a symmetric interaction between the system and a macroscopic system which acts as a program for the desired  transformation.
Their work provides an example of the trend  of applying the formalism of generalized probabilistic theories to a broader spectrum of topics in physics.

Fuchs and Stacey (chapter 8) provide some critical remarks on existing axiomatizations of quantum theory,  and express some desiderata for future work.  
 In addition to motivating the search for a more compelling picture, they review
   the QBist approach to the foundations of quantum theory \cite{qubist0,qubist1,qubist2,qubist3},
which aims to understand quantum theory within a subjective Bayesian approach to probability theory,
in particular, as a modification to 
 the manner in which experimental probabilities in different counterfactual scenarios are related to one another.     
 
    

 
\medskip 

\subsection{Categories and convex sets}


Chapters 9 and 10 expound the foundations of the category-theoretic framework and of the framework of convex operational theories, respectively. 
As we have noted, 
developing suitable frameworks is an essential step  in the axiomatization of quantum theory and a subject of active research in its own right.  
The reader may well wonder why we chose to put the framework chapters \emph{after} the axiomatizations chapters, rather than before.    
  There are several reasons for our choice. First of all,  the main message of the axiomatizations can be easily grasped without  entering into the specific details of the framework.  In fact, given the richness of nuances contained in the axiomatization works, too much attention to details could even hinder the first reading.  
 On top of that, the frameworks used in the axiomatization chapters are often different from those presented in chapters 9 and 10.  Finally, giving first a taste of what the study of operational theories 
 can achieve is probably the best way to motivate the reader to a deeper excursion into the structural aspects of the  framework.     

Our excursion starts with 
Coecke, Duncan, Kissinger, and Wang (chapter 9), who review the category-theoretic framework for describing  operational theories \cite{abracoecke,cqm,guises,contemporarybob}. 
This chapter will take the reader through the 
quantum structures
  that are central to this approach, such as the tensor product structure, the compact structure associated to 
quantum teleportation, the dagger structure associated to the adjoint, and the Frobenius structure associated to orthonormal bases. 
These notions are expressed in terms of a diagrammatic calculus that allows mathematical proofs to be carried out entirely through the manipulation of diagrams.
Using this framework,  the authors provide  a purely graphical treatment  of complementarity and of the Greenberger-Horne-Zeilinger paradox at the end of the chapter.

 Barnum and Wilce (chapter 10)  present  the framework of convex operational theories   \cite{hardy2001,barrett,nobroad,telep,barnumwilce}. 
   Here, the  structures of ordered vector spaces,  geometry, and symmetry are the main protagonists.  States of a given system are represented by points in a finite dimensional convex set, measurements by positive linear functionals, and physical transformations by positive linear maps.  To illustrate some of these notions, the chapter presents many concrete examples of convex operational theories that are nonclassical but distinct from quantum theory. The treatment of tensor products and entanglement in convex operational theories is reviewed, as is the question of which information processing advantages of quantum theory are generic to convex operational theories that are nonclassical.  Finally, the chapter discusses axioms for quantum theory based on considerations of symmetry and composition. 
  

\medskip  

\subsection{Quantum versus super-quantum correlations}


In the final part of this book we present a number of important features of the set of quantum correlations.   

Pawlowski and Scarani  (chapter 11) discuss the principle of Information Causality \cite{infocau}.  
 This is a device-independent principle that concerns the possibilities for communication within an operational theory, in particular, for a communication protocol known as a random access code, wherein the receiver only gets part of the data encoded by the sender, but is allowed to choose which part.  An operational theory is said to be information causal if assisting a random access code with an arbitrary shared nonsignalling resource of correlations provides no advantage. 
 They show that Information Causality  
 implies the Tsirelson bound and many other features of the set of quantum correlations.  

Navascu\'es (chapter 12)  discusses the principle of Macroscopic Locality \cite{macroloc}.   This approach makes use of the fact that the strength of nonsignalling correlations among microscopic systems has consequences for the strength of such correlations among macroscopic systems, that is, among collections of microscopic systems wherein one cannot address the constituents individually.  To insist that an operational theory satisfy macroscopic locality is to insist that  in the macroscopic limit it must look classical, in particular, it must look local in the sense of not violating a Bell inequality.   This principle implies that the microscopic correlations must satisfy the Tsirelson bound and reproduces other features of the quantum set.


Ac\'in,   Almeida,   Augusiak, and  Brunner (chapter 13) describe the foundational implications of a multipartite game called Guess Your Neighbor's Input (GYNI) \cite{gyni}. The game of GYNI is one  for which quantum does {\em not} provide an advantage over classical, but for which nonsignalling alternatives to quantum theory do provide an advantage. Thus, the game provides a natural separation between quantum correlations and superquantum correlations.  Various consequences for the project of deriving the quantum set are discussed:
GYNI can be used to show that 
in the multipartite scenario, the no-signalling principle and the assumption that systems locally look quantum is not enough to recover the quantum set, unlike the bipartite case; and to derive Bell inequalities that do not admit of any quantum violation.

Finally, Colbeck and Renner (chapter 14)  consider the question of whether there might exist an extension of quantum theory, that is, an alternative theory that enables predictions that have less uncertainty than those of quantum theory, but which reproduces the quantum predictions when one averages over certain variables \cite{noext1}.  Using an assumption that seeks to formalize the notion that observers are free to choose the settings of their measurements, they prove a result that rules out such extensions.  




\section{Concluding Remarks}


The goal of understanding what physical principles might underlie the formalism of quantum theory is an ambitious one. 
Nonetheless, this monograph testifies to the fact that real and sustained progress on the question has been achieved in recent years.   We hope that readers will come away with a sense of the excitement and promise of contemporary research in the field of quantum foundations and that some may be inspired to contribute to the endeavour themselves in the years to come.

\acknowledgements 

The authors acknowledge the support of the Perimeter Institute for Theoretical Physics during the course of this project.   Research at Perimeter Institute is supported by the Government of Canada through Industry Canada and by the Province of Ontario through the Ministry of Research and Innovation.  GC also acknowledges the Foundational Questions Institute (FQXi-RFP3-1325), the National Natural Science Foundation of China (Grants 11450110096 and 11350110207), and  the 1000 Youth Fellowship Program of China.

\end{document}